\pdfoutput=1

\documentclass[11pt]{article}

\usepackage[dvipsnames]{xcolor}
\usepackage{ACL2023}

\usepackage{times}
\usepackage{latexsym}

\usepackage[T1]{fontenc}

\usepackage[utf8]{inputenc}

\usepackage{microtype}

\usepackage{subcaption}
\usepackage{siunitx}
\usepackage{inconsolata}
\usepackage{amsmath}
\usepackage{graphicx}
\usepackage{multirow}
\usepackage{booktabs}
\usepackage{bbding}
\usepackage{MnSymbol}
\usepackage{enumitem}
\usepackage{svg}
\usepackage{pifont}
\usepackage[normalem]{ulem}
\usepackage{enumitem}

\newcommand{\xmark}{\ding{55}}%
\newcommand\ILPCR{\textbf{\texttt{IL-PCR}}}
\title{U-CREAT: Unsupervised Case Retrieval using Events extrAcTion }

\author{Abhinav Joshi\thanks{\ \ Equal Contributions} \qquad Akshat Sharma\footnotemark[1] \qquad Sai Kiran Tanikella\footnotemark[1] \qquad Ashutosh Modi \\
Indian Institute of Technology Kanpur (IIT Kanpur)\\
\texttt{\{ajoshi, akshatsh, tskiran, ashutoshm\}@cse.iitk.ac.in} 
}

\begin{document}
\maketitle

\begin{abstract}

The task of Prior Case Retrieval (PCR) in the legal domain is about automatically citing relevant (based on facts and precedence) prior legal cases in a given query case. To further promote research in PCR, in this paper, we propose a new large benchmark (in English) for the PCR task: IL-PCR (Indian Legal Prior Case Retrieval) corpus. Given the complex nature of case relevance and the long size of legal documents, BM25 remains a strong baseline for ranking the cited prior documents. In this work, we explore the role of events in legal case retrieval and propose an unsupervised retrieval method-based pipeline U-CREAT (Unsupervised Case Retrieval using Events Extraction). We find that the proposed unsupervised retrieval method significantly increases performance compared to BM25 and makes retrieval faster by a considerable margin, making it applicable to real-time case retrieval systems. Our proposed system is generic, we show that it generalizes across two different legal systems (Indian and Canadian), and it shows state-of-the-art performance on the benchmarks for both the legal systems (IL-PCR and COLIEE corpora). 



\end{abstract}
\section{Introduction} \label{sec:intro}

Traditionally, in the legal domain, for a given legal case (query document) at hand, lawyers and judges have relied on their expertise and experience to cite relevant past precedents (cited documents). Moreover, even when legal professionals have made limited use of technology, it has been mainly restricted to Boolean queries and keywords. However, as cases increase, it becomes difficult for even experienced legal professionals to cite older precedents. NLP-based technologies can aid legal professionals in this regard. The task of \textit{Prior Case Retrieval} (PCR) has been formulated to address this problem \cite{COLIEE2021}. More concretely, the task of \textit{Prior Case Retrieval} involves retrieving all the previous legal documents that should be cited in the current legal document based on factual and precedent relevance. PCR can be particularly important in populous countries like India, where the number of cases has been growing exponentially, for example, there are 41 million pending cases in India \cite{njdc-district}. Technology-based solutions such as PCR can make the process streamlined and efficient, expediting case disposal.

\begin{figure}[t]
\centering
  \includegraphics[scale=0.19]{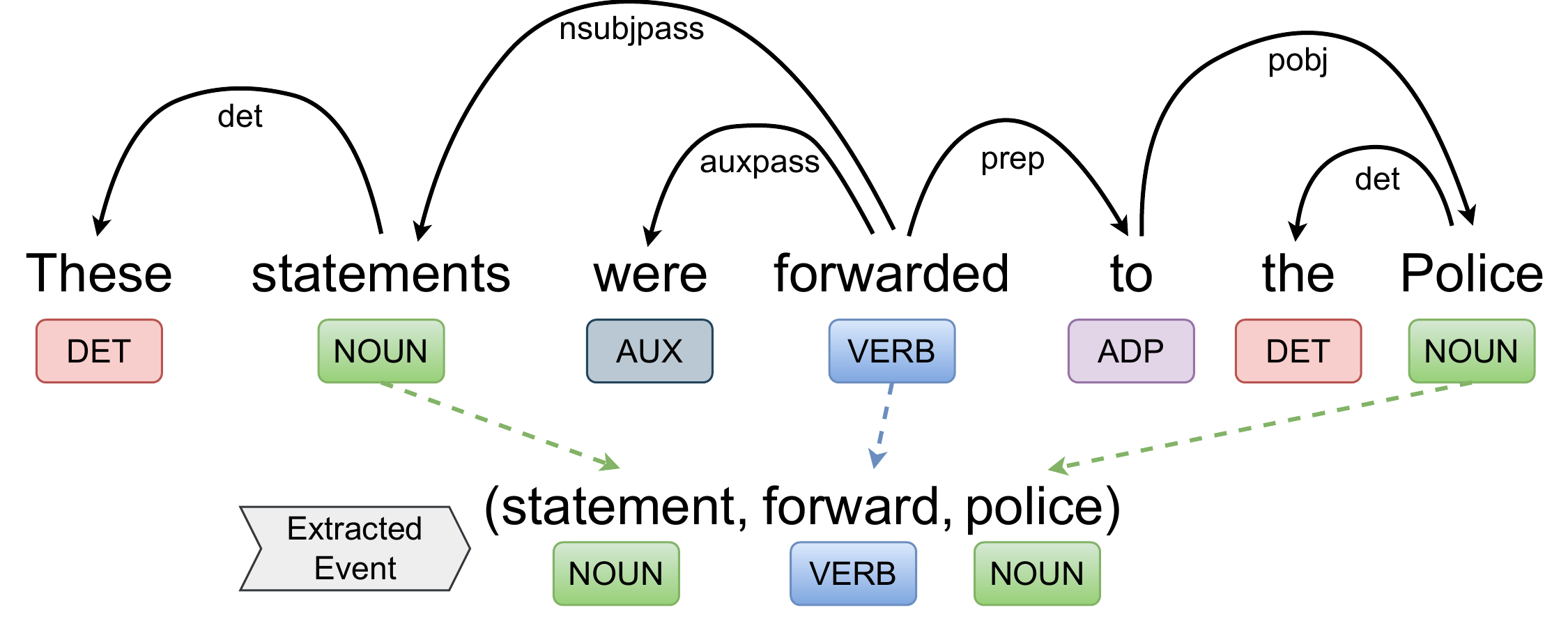}
  \caption{Dependency parse of the sentence (along with extracted event) from the \ILPCR\ corpus: ``These statements were forwarded to the Police".}
  \label{fig:dependency_parsing_example}
\end{figure}

\begin{figure}[t]
\centering
  \includegraphics[scale=0.29]{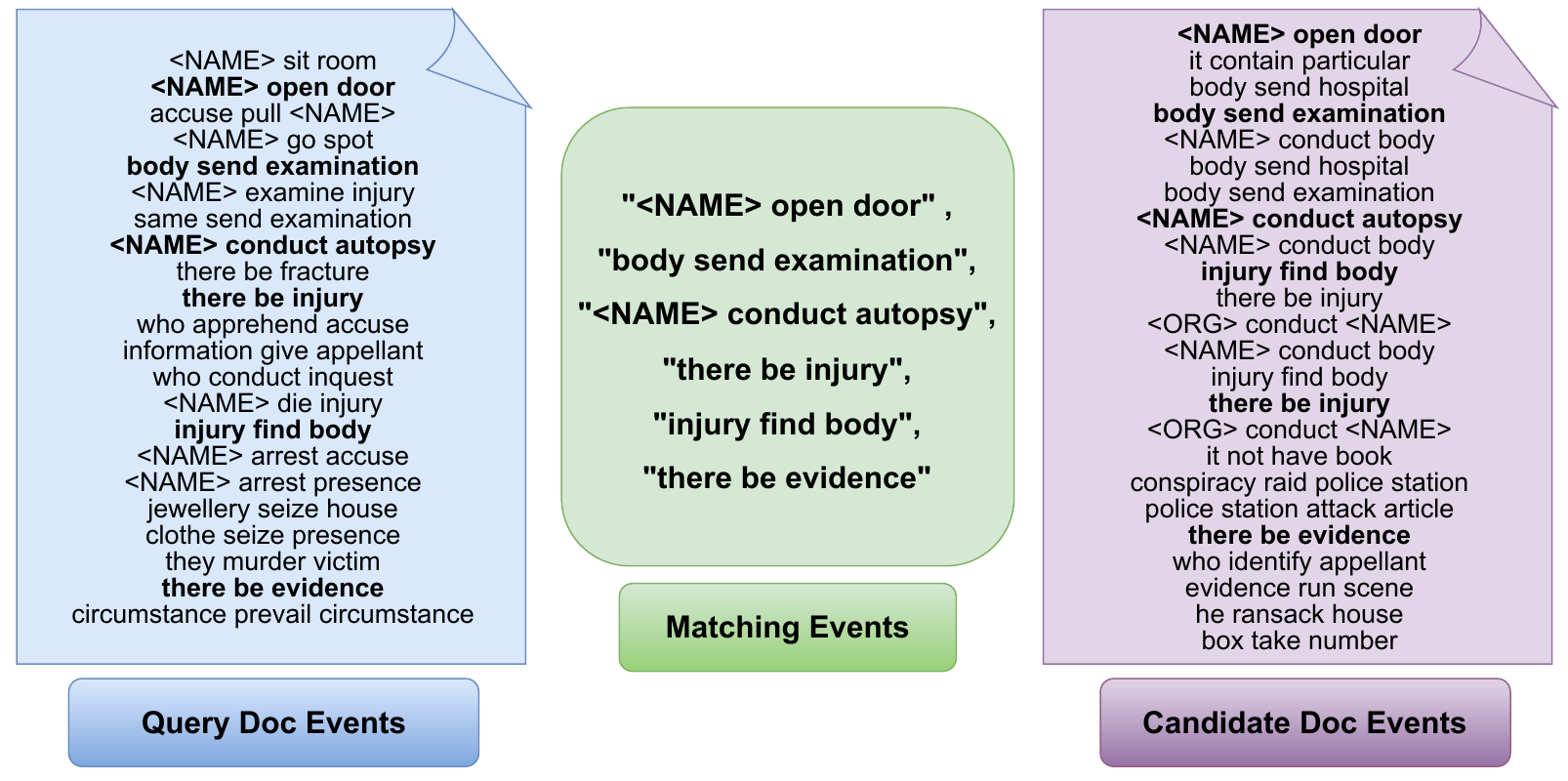}
  \caption{The Figure shows common events (highlighted in \textbf{bold}) for a positive query-candidate pair (example taken from  the \ILPCR\ corpus)}
  \label{fig:event_query_candidate_example}
\end{figure}

\noindent PCR is different from standard document retrieval tasks. It is primarily due to the nature of legal documents themselves. Legal documents, in general, are quite long (tens to hundreds of pages), which makes each document in both the query and candidate pool long. Legal documents are unstructured and sometimes noisy (for example, in many common law countries like India, legal documents are manually typed and prone to grammatical and spelling mistakes). Moreover, in a common-law system, where the judges can overrule an existing precedence, there is some degree of subjectivity involved,   making the task of document processing and retrieval challenging. 

\noindent In this paper, we propose a new large PCR corpus for the Indian legal setting referred to as Indian Legal Prior Case Retrieval (\ILPCR) corpus. Further, we propose an unsupervised approach for the task of prior case retrieval based on events structure in the document. Events are defined in terms of predicate and its corresponding arguments (see Figure   \ref{fig:dependency_parsing_example}) obtained via a syntactic dependency parser.  The proposed event-based representation technique performs better than the existing state-of-the-art approaches both in terms of retrieval efficiency as well as inference time. We conjecture that events obtained via a dependency parser play an essential role in providing a short summary of long judgment documents, hence reducing the noise (task-dependent non-relevant information) by a considerable margin (also shown in Fig.  \ref{fig:event_query_candidate_example}).

\noindent The focus of this paper is an unsupervised and fast approach for retrieving relevant legal documents, in contrast to resource and compute-intensive supervised approaches. In the legal domain, supervised algorithms often require hand-crafted engineering/tuning with considerable experimentation to enable deployment in a real-time scenario, making them harder to adapt to an industrial setting.  
Although not a fair comparison, our proposed method shows an improvement of $5.27$ F1 score over a recent state-of-the-art supervised method \cite{ECIR-22-bert} for the existing PCR benchmark dataset of COLIEE'21  (\S\ref{sec:existing-results}). In a nutshell, we make the following contributions:
\begin{itemize}[noitemsep,nosep]
\item Considering the lack of available benchmarks for the Indian legal setting, we create a new benchmark for Prior Case Retrieval focused on the Indian legal system (\ILPCR) and provide a detailed analysis of the created benchmark. Due to the large size of the corpus, the created benchmark could serve as a helpful resource for building information retrieval systems for legal documents (\S\ref{sec:corpus}). We release the corpus and model code for the purpose of research usage via GitHub: \url{https://github.com/Exploration-Lab/IL-PCR}.  

\item We propose a new framework for legal document retrieval: U-CREAT (Unsupervised Case Retrieval using Events Extraction), based on the events extracted from documents. We propose different event-based models for the PCR task. We show that these perform better than existing state-of-the-art methods both in terms of retrieval efficiency as well as inference time (\S\ref{sec:all-results}). 

\item Further, we show that the proposed event-based framework and models generalize well across different legal systems (Indian and Canadian systems) without any law/demography-specific tuning of models. 
\end{itemize} 

\section{Related Work} \label{sec:relatedwork}

Automating processes and tasks in the legal domain has been an active area of research in the NLP and IR community in the past few years. For example, several tasks/research problems and solutions have been proposed, e.g., Catchphrase Extraction \cite{Galgani2012}, Crime Classification \cite{Wang2018,Wang2019}, Summarization \cite{Tran2019}, Rhetorical Role prediction \cite{malik-rr-2021,kalamkar-etal-2022-corpus} and Judgment Prediction \cite{LJPAAAI2020,malik2021ildc,chalkidis-etal-2019-neural,Aletras2016,Chen2019,LongT2019,Xu2020,Yang2019,kapoor-etal-2022-hldc}. 

Some earlier works \cite{ML-PCR,Jackson2003} in Prior Case Retrieval have used feature-based machine learning models such as SVM. Since the past few years, the Competition on Legal Information Extraction and Entailment (COLIEE) has been organized annually \cite{COLIEE2021}. COLIEE has spurred research in PCR. Researchers participating at COLIEE have shown that BM-25 based method is a strong baseline. Most of the participating systems in COLIEE have used models based on BM-25 combined with other techniques like TF-IDF, language models, transformers, and XG-Boost (e.g., \cite{Yes-BM25,COLIEE2021,DESIRES,Ma2021RetrievingLC,JNLP,Shao2020BERTPLIMP,UIRPC}). Citation network-based approaches \cite{PB1,PB6,PB,PB3,PB4} are not meaningfully applicable to PCR as the legal citation networks are quite sparse. \citet{ECIR-22-bert} proposed BERT-based Query-by-Document Retrieval method with Multi-Task Optimization. We also experimented with transformer-based methods for retrieving prior cases as described in \S\ref{sec:baselines}. 

\noindent In the NLP community, researchers have used event-based information for many different Natural Language Understanding (NLU),  and commonsense reasoning tasks \cite{chen-etal-2021-event, Event_Chains,Event_chains_participants,modi-titov-2014-inducing,modi-2016-event,modi-etal-2017-modeling}. For example, \citet{glavavs2014event} extracted events from a document and used the event-centric graph representation for information retrieval and multi-document summarization tasks, where they define an event as a tuple of predicate (action) and corresponding arguments (participants/actors). In the legal-NLP domain, event-based representations have not been explored much, as also pointed out in the survey by \citet{feng2022legal}. In this work, we employ event-based representation for PCR. 

\section{\textbf{\ILPCR}\  Corpus and PCR Task} \label{sec:corpus}

To spur research in the area of PCR, we propose the creation of a new corpus for the task of PCR: Indian Legal Prior Case Retrieval (\ILPCR) corpus. \ILPCR\ corpus is a corpus of Indian legal documents in English containing 7070 legal documents. 

\subsection{\ILPCR\ Corpus Creation}
The corpus is created by scraping legal judgment documents (in the public domain) from the website IndianKanoon (\url{https://indiankanoon.org/}). We 
started by scraping documents corresponding to the top $100$ most cited Supreme Court of India (SCI) cases (these are termed the zero-hop set). To gather more cases, we scraped documents cited within the zero-hop cases to obtain the one-hop cases. Scraping in this manner ensured a sufficient number of cited cases for each document. In practice, gathering cases till the second hop was sufficient for a corpus of desirable size. The desirable size is decided by comparing it relatively to the size of the existing PCR benchmarks like COLIEE. Any empty/non-existent cases were removed. Zero and one-hop cases were merged into a large query pool, which was further split into the train ($70\%$), validation ($10\%$), and test ($20\%$) queries. To facilitate generalization among developed models, we did not put any temporal constraints on the scraped documents (as also justified in \cite{malik2021ildc}); the cases range from 1950 to 2020. We followed a similar corpus creation methodology as done by the COLIEE benchmark. 

\noindent\textbf{Pre-Processing:}  All documents are normalized for names and organization names using a NER model \cite{Spacy-io} and a manually compiled gazetteer. This step helps to create more generic event representations. As done in the case of other PCR corpora such as COLIEE \cite{COLIEE2021}, the text segment associated with each citation (these are in the form of hyperlinks in scraped documents) is replaced with a citation marker \texttt{<CITATION>}. The text segments corresponding to statutes (acts and laws) are not replaced since our focus is prior case retrieval and not statute retrieval \cite{kim2019statute}. We also experimented with another version of the corpus where the entire sentence containing the citation is removed (details in \S\ref{sec:analysis}). 

\begin{table}[t]
\small
\renewcommand{\arraystretch}{0.7}
\setlength\tabcolsep{2pt}
\hspace{0.2cm}
\centering
\begin{tabular}{ccc}
\toprule
\textbf{Dataset}              & \textbf{COLIEE'21} & \ILPCR \\ \midrule
\# Documents                  &   4415              &   7070                  \\
Avg. Document Size            &    5813.66             &        8093.19             \\
\# query Documents            &         900        &         1182            \\
Vocab Size                    &         80577        &    113340                 \\
Total Citation Links          &        4211         &       8008              \\
Avg. Citation Links per query &         4.678        &      6.775               \\ 
Language &         English        &      English              \\
Legal System &         Canadian        &      Indian              \\
\bottomrule
\end{tabular}
\caption{The table compares the created \ILPCR\  corpus with COLIEE'21 corpus.}
\label{tab:il-pcr-coliee}
\vspace{-3mm}
\end{table}

\noindent\textbf{Comparison with Existing Corpora:} 
We compare existing PCR corpus from COLEE'21 and \ILPCR\ in Table \ref{tab:il-pcr-coliee}. \ILPCR\  is almost $1.6$ times COLIEE 2021 and average length of document in \ILPCR\ is almost $1.4$ times. \ILPCR\  has a much larger vocabulary and more citations per document. Both COLIEE 2021 and \ILPCR\ are primarily in English but address different legal systems, namely, Canadian and Indian legal systems respectively. 

\subsection{PCR Task Definition}
Given a legal document as a query $\mathcal{Q}_i$ and a pool of $N$ legal documents as candidates $\{ \mathcal{C}_1, \mathcal{C}_2., \ldots, \mathcal{C}_N \}$, the Prior Case Retrieval task is to retrieve the legal documents from the candidate pool which are relevant (and hence cited) in the given query document. As also pointed out by the legal expert, relevance in the legal domain is mainly about similar factual situations and previous legal precedents. 

\section{Event Based Representations} \label{sec:events}

A story or an incident is best described in terms of a sequence of events \cite{Event_Chains,Event_chains_participants,chen-etal-2021-event}. If we consider a case judgment document to be a narrative about how things (e.g., situations in the form of facts) developed, then it is best to represent a legal document in terms of  events. We define an event as a tuple containing predicate (describing the main action, typically it is verb/verb-compound) and its main arguments (describing main actors/participants, typically these correspond to subject, object, indirect object, and prepositional object) as shown in Fig. \ref{fig:dependency_parsing_example}.

\subsection{Event Extraction}
To extract events, legal documents are first pre-processed to remove noise (unwanted characters and symbols) using regex-based patterns. For example, initials (not picked by NER) in the names (e.g., initials A.R. in the name  A. R. Lakshman) are removed. Similarly, characters other than letters and citation markers are removed. Honorifics like Dr., Mr., Mrs., etc. are removed as these were wrongly picked up as the end of the sentence during sentence splitting and during event extraction. Other short forms like no., nos., addl., etc., are replaced by corresponding full words. Subsequently, a dependency parser is used to extract events from texts. 

A dependency parser represents a sentence in the form of a directed graph $G: (V, E)$, where $V$ are vertices representing words and $E$ are the directed edges that capture the grammatical (syntactic) relationship between words \cite{kubler2009dependency}. Sentences in the document are parsed with the dependency parser (we use spaCy: \cite{Spacy-io}) to extract the list of verbs. These verbs form the root of the dependency graph. As observed, mostly the sentences in legal documents are in active voice. The left children of each verb are examined to find the subjects with syntactic dependency relationships like nsubj, nsubjpass, and csubj. The right children of a verb are considered for relationships like dobj, pobj, and dative to indicate the object's presence. Further, the lefts and rights are examined for conjunctions and compounds to get all the possible subjects and (indirect) objects. Each of the words in the extracted event is lemmatized to make the event more generic. Further, incomplete events and empty events (generated due to incorrect sentence splitting) are discarded. Both query and candidate documents are processed with the dependency parser to get the events. After removing noisy events, we did not observe any significant mistakes in the extracted events. Manual examination of the verb-argument tuples showed plausible events. 

Events play an important role in establishing the relationship between a case and a cited (precedent) case. If a case has a precedent, then most likely, both are related based on the nature of the facts, evidence, and judgment. The events in a prior case form a basis for the arguments and judgments in such similar cases. Based on the experimental results, we conjecture that events further help to summarize documents in terms of main actions (e.g., related to facts) and hence help to filter out noise. 

\section{Experiments, Results and Analysis} \label{sec:all-results}

\noindent\textbf{Datasets:} We experimented with the COLIEE-21 and \ILPCR\ corpora. Since the two corpora are different, it enables checking the generalization capabilities of models. 

\noindent\textbf{Evaluation Metric:} We use a micro-averaged F1 score as the evaluation metric (as done in COLIEE-21\footnote{Section 3.1 in \url{https://sites.ualberta.ca/~rabelo/COLIEE2021/}}). In practice, models predict a relevance score for each candidate for a given query. Top-K-ranked candidates are considered for prediction (i.e., whether a candidate is cited or not). As done in previous work \cite{COLIEE2021}, we select K based on the best performance on the validation set and report the F1 on the test set using the same K value (metric definition is provided in App. \ref{app:eval-metric}). 

\begin{figure*}[t]
\centering
  \includegraphics[scale=0.35]{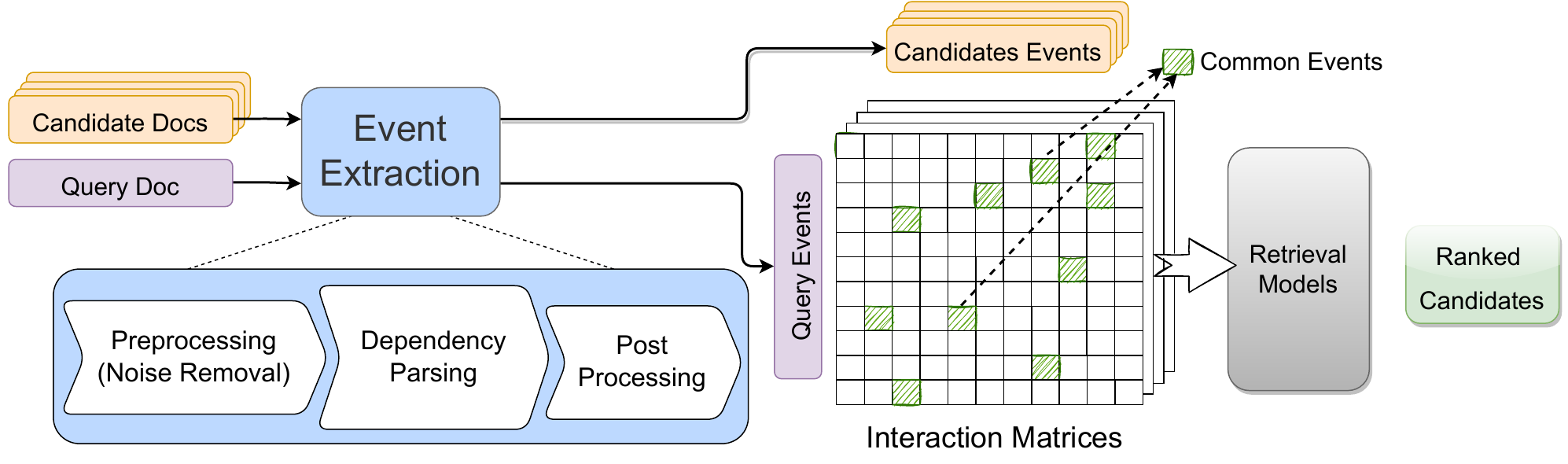}
  \caption{U-CREAT pipeline based on events extraction, for the PCR task.}
  \label{fig:U-CREATE_pipeline}
  \vspace{-5mm}
\end{figure*}

\subsection{Baselines, Proposed Models and Results} \label{sec:baselines}

For the baseline models, we selected the prominent approaches used for the PCR task. Considering the findings reported in COLIEE-21, BM25 marks a strong baseline \cite{Yes-BM25,COLIEE2021} for document retrieval tasks in the legal domain. Moreover, most of the re-ranking-based supervised methods \cite{DESIRES,JNLP,UIRPC,Shao2020BERTPLIMP,ECIR-22-bert} also use BM25 as a pre-filtering step for document retrieval. Broadly, we consider three types of unsupervised retrieval models as baselines, 1) Word-based (Count-based), which are lexical models using words directly; 2) Transformer-based models, which  capture the semantics using distributed representations of words; and 3) Sentence Transformer based models, which capture semantics at the sentence level. We provide experimental results for all the baseline models on COLIEE'21 and \ILPCR\ datasets in Table \ref{tab:experiment_results}. We describe baseline models next. 

\noindent\textbf{Word-Based (Count-Based):} We use a standard implementation of BM25 (Sklearn's \cite{scikit-learn} TfidfVectorizer module) to compute scores for each query-candidate pair. We experiment with two widely used versions of BM25, unigram, and bigram. The bigram variant of BM25 improves the retrieval performance (Table \ref{tab:experiment_results}) by a considerable margin, from $14.72$\% to $22.14$\% in COLIEE'21 and $13.85$\% to $28.59$\% in \ILPCR. However, the large runtime overhead of the bigram setting makes it ineffective for a real-time retrieval system and hence is usually not the preferred choice. 

\noindent\textbf{Transformer-Based:}  We use two widely used transformer models for generating word embedding: pre-trained BERT \cite{devlin-etal-2019-bert} and DistilBERT \cite{sanh2019distilbert}. We also experiment with a fine-tuned version. We fine-tune the model on the train split of the respective datasets (\ILPCR\ and COLIEE'21) using standard masked language modeling (MLM) objective (details in App.\! \ref{app:hyper-params}). In addition, we also experiment with Indian legal domain-specific language models: InCaseLawBERT and InLegalBERT \cite{InCaseLawBERT}. 
We use transformer models in two settings, one using the entire document and the other using the top 512 tokens. Due to limitations on the input size of transformer models, to learn the representation of the entire document, we divide the document into multiple segments (each of 10 sentences) with a stride of 5 sentences (to ensure overlap). Subsequently, an interaction matrix (having relevance score) between query and candidate segments is created using cosine similarity between respective representations and this is followed by an aggregation step (avg. or max) to come up with a score. In the other setting, we consider only the top 512 tokens as input to the transformer and discard the remaining information. Our experiments highlight that fine-tuning these models slightly improves the performance in the case of transformers with top 512 tokens and slightly worsens the performance in the case of full document transformers. (Table \ref{tab:experiment_results}). We observe that InCaseLawBERT and InLegalBERT perform quite poorly, possibly due to noise in legal documents.

\noindent\textbf{Sentence Transformer-Based (SBERT):}
We also experiment with sentence embeddings-based methods that capture the similarity at the sentence level. We experiment with two popularly used sentence embedding methods\footnote{For model implementation, we used the SBERT  library (\url{https://www.sbert.net/examples/unsupervised_learning/SimCSE/README.html}). We used the hyper-parameters corresponding to the best-performing model on the leaderboard for the sentence similarity task.}: SBERT-BERT and SBERT-DistilRoBERTA \cite{reimers2019sentence}. To finetune the transformers in an unsupervised fashion, we follow SimCSE's \cite{simcse} strategy (details in App. \ref{app:hyper-params} and App. \ref{app:simcse}). For all the methods, we use cosine similarity between all query-candidate sentence pairs to generate an interaction matrix and consider the max of the matrix to be the relevance score for the pair. In general, compared to full document and vanilla transformers SBERT based approaches have better performance (Table \ref{tab:experiment_results}). 

\noindent \textbf{Event Based Models:} The general pipeline for event-based models is shown in Fig. \ref{fig:U-CREATE_pipeline}. We refer to this pipeline as U-CREAT (Unsupervised Case Retrieval using Event extrAcTion). We first extract event representations from the query and candidate documents, and these are used to calculate an interaction matrix between each query-candidate pair. The interaction matrix captures similarities between events (relevance scores); subsequently, a retrieval model is used to rank the candidates. The methods proposed below differ in the document representation, interaction matrix, and retrieval model.  

\noindent\textbf{Atomic Events:} In this variant, an event (predicate and arguments tuple) is considered as an atomic unit (like a word), and a document is represented only by these atomic events. An approach to generating the relevance scores can be using Jaccard similarity (IOU: Intersection Over Union) over the obtained set of events. For a given query candidate pair $(\mathcal{Q}_i, \mathcal{C}_j)$, we extract the events corresponding to each document, 
$\mathcal{E}^{(\mathcal{Q}_i)} =\{e^{(\mathcal{Q}_i)}_1,  \ldots,  e^{(\mathcal{Q}_i)}_n\}$, and 
$\mathcal{E}^{(\mathcal{C}_j)} = \{e^{(\mathcal{C}_j)}_1,  \ldots,  e^{(\mathcal{C}_j)}_m\}$ which is used to compute the Jaccard similarity, i.e., 
$\text{Relevance Score} = \frac{|\mathcal{E}^{(\mathcal{Q}_i)} \cap \mathcal{E}^{(\mathcal{C}_j)}|}{|\mathcal{E}^{(\mathcal{Q}_i)} \cup \mathcal{E}^{(\mathcal{C}_j)}|}$. 
As shown in Table \ref{tab:experiment_results}, this trivial strategy of computing Jaccard similarity over the set of events improves performance on both datasets compared to BM25. Though the gain is less in COLIEE'21 (increase by $\sim 8$ F1 score ), in \ILPCR, the improvement is significant (increase by $\sim 20$ F1 score). We speculate that given the legal document's diverse and lengthy nature, events help filter out the noise and improve performance significantly. Another way of getting the relevance score would be to take all the extracted events $\mathcal{E}^{(\mathcal{Q}_i)}$ and $\mathcal{E}^{(\mathcal{C}_j)} $ and perform a BM25 over atomic events instead of words; this setting helps to capture the relation between various events present in both the docs. We experiment with multiple settings of BM25.  
The results highlight that the BM25's unigram setting performs similarly to the Jaccard similarity with a drop in performance when increased to bigram, trigram, the reason being the lower frequency of bigram/trigram events present in the document pairs. 

\begin{table*}[h]
\small
\renewcommand{\arraystretch}{1}
\setlength\tabcolsep{11pt}
\centering
\begin{tabular}{cccc}
\toprule 
\multicolumn{2}{c}{\textbf{Model}}                                                      & \textbf{COLIEE'21} & \textbf{\ILPCR} \\ \midrule 
\multirow{2}{*}{Word Level (Count Based)}               & BM25                              & 14.72 \textbf{(Baseline)}              & 13.85 \textbf{(Baseline)}              \\
                                          & BM25 (Bigram)                     & 22.14 (\textcolor{ForestGreen}{ $\uparrow$ 7.42})      & 28.59 (\textcolor{ForestGreen}{ $\uparrow$ 14.74})      \\
                                          \midrule
\multirow{6}{*}{\begin{tabular}{@{}c@{}}Segmented-Doc  \\ Transformer \\(full document) \end{tabular}}     
                                          & BERT                              & 5.10 (\textcolor{red}{ $\downarrow$ 9.62})             & 9.24 (\textcolor{red}{ $\downarrow$ 4.61})              \\
                                          & BERT (finetuned)                  & 4.58 (\textcolor{red}{$\downarrow$ 10.14})              & 7.91 (\textcolor{red}{ $\downarrow$ 5.94})       \\
                                          & DistilBERT                        & 10.04 (\textcolor{red}{ $\downarrow$ 4.68})              & 16.61 (\textcolor{ForestGreen}{ $\uparrow$ 2.76})         \\
                                          & DistilBERT (finetuned)            & 4.73 (\textcolor{red}{ $\downarrow$ 9.99})              & 7.86 (\textcolor{red}{ $\downarrow$ 5.99})         \\
                                          & InCaseLawBERT                     & 1.71 (\textcolor{red}{ $\downarrow$ 13.01})             & 3.62 (\textcolor{red}{ $\downarrow$ 10.23})                 \\
                                          & InLegalBERT                       & 2.79 (\textcolor{red}{ $\downarrow$ 11.93})             & 7.57 (\textcolor{red}{ $\downarrow$ 6.28})                   \\
                                          \midrule
\multirow{6}{*}{\begin{tabular}{@{}c@{}} Transformer \\(top 512 tokens) \end{tabular}}
                                          & BERT                              & 0.53 (\textcolor{red}{ $\downarrow$ 14.19})             & 0.56 (\textcolor{red}{ $\downarrow$ 13.29})               \\
                                          & BERT (finetuned)                  & 0.46 (\textcolor{red}{ $\downarrow$ 14.26})             & 0.88 (\textcolor{red}{ $\downarrow$ 12.97})               \\
                                          & DistilBERT                        & 0.54 (\textcolor{red}{ $\downarrow$ 14.18})             & 0.50 (\textcolor{red}{ $\downarrow$ 13.35})               \\
                                          & DistilBERT (finetuned)            & 0.34 (\textcolor{red}{ $\downarrow$ 14.38})             & 0.75 (\textcolor{red}{ $\downarrow$ 13.1})                \\
                                          & InCaseLawBERT                     & 0.78 (\textcolor{red}{ $\downarrow$ 13.94})             & 0.75 (\textcolor{red}{ $\downarrow$ 13.1})                \\
                                          & InLegalBERT                       & 0.50 (\textcolor{red}{ $\downarrow$ 14.22})             & 0.71 (\textcolor{red}{ $\downarrow$ 13.14})                 \\
                                          \midrule
\multirow{4}{*}{\begin{tabular}{@{}c@{}} Sentence \\ Transformer \\(SBERT) \end{tabular}}
                                          & BERT                              & 6.79 (\textcolor{red}{ $\downarrow$ 7.93})              & 5.94 (\textcolor{red}{ $\downarrow$ 7.91})                  \\
                                          & DistilRoBERTa                     & 3.63 (\textcolor{red}{ $\downarrow$ 11.09})              & 3.91 (\textcolor{red}{ $\downarrow$ 9.94})                   \\
                                          & BERT (finetuned)                  & 7.68 (\textcolor{red}{ $\downarrow$ 7.04})              & 6.01 (\textcolor{red}{ $\downarrow$ 7.84})                     \\
                                          & DistilRoBERTa (finetuned)         & 1.26 (\textcolor{red}{ $\downarrow$ 13.46})              & 2.14 (\textcolor{red}{ $\downarrow$ 11.17})\\
                                          \hline \hline
\multirow{4}{*}{Atomic Events}        & Jaccard similarity                               & 23.08 (\textcolor{ForestGreen}{ $\uparrow$ 8.36})               & 34.17 (\textcolor{ForestGreen}{ $\uparrow$ 20.32})     \\
                                        & BM25                              & 23.45 (\textcolor{ForestGreen}{ $\uparrow$ 8.73})               & 36.77 (\textcolor{ForestGreen}{ $\uparrow$ 22.92})               \\
                                          & BM25 (Bigram)                     & 22.42 (\textcolor{ForestGreen}{ $\uparrow$ 7.70})      & 31.81 (\textcolor{ForestGreen}{ $\uparrow$ 17.96})               \\
                                          & BM25 (Trigram)                    & 21.12 (\textcolor{ForestGreen}{ $\uparrow$ 6.40})               & 27.61 (\textcolor{ForestGreen}{ $\uparrow$ 13.76})               \\
                                          \midrule
                                          
\multirow{5}{*}{Non-atomic Events}        & BM25                              & 14.19 (\textcolor{ForestGreen}{ $\uparrow$ 0.53})                       & 11.99 (\textcolor{red}{ $\downarrow$ 1.86})               \\
                                          & BM25 (Bigram)                     & 23.59 (\textcolor{ForestGreen}{ $\uparrow$ 8.87})         & 32.27 (\textcolor{ForestGreen}{ $\uparrow$ 18.42})               \\
                                          & BM25 (Trigram)                    & 24.13 (\textcolor{ForestGreen}{ $\uparrow$ 9.41})              & 36.53 (\textcolor{ForestGreen}{ $\uparrow$ 22.68})      \\
                                          & BM25 (Quad-gram)                  & 22.69 (\textcolor{ForestGreen}{ $\uparrow$ 7.97})               & 34.76 (\textcolor{ForestGreen}{ $\uparrow$ 20.91})               \\
                                          & BM25 (Penta-gram)                 & 21.81 (\textcolor{ForestGreen}{ $\uparrow$ 7.09})               & 33.54 (\textcolor{ForestGreen}{ $\uparrow$ 19.69})               \\
                                          \midrule

\multirow{5}{*}{Events Filtered Docs}      & BM25                              & 18.97 (\textcolor{ForestGreen}{ $\uparrow$ 4.25})      & 19.64 (\textcolor{ForestGreen}{ $\uparrow$ 5.79})               \\
                                          & BM25 (Bigram)                     & 23.3 (\textcolor{ForestGreen}{ $\uparrow$ 8.58})               & 30.28 (\textcolor{ForestGreen}{ $\uparrow$ 16.43})               \\
                                          & BM25 (Trigram)                    & \textbf{27.32} (\textcolor{ForestGreen}{ $\uparrow$ \textbf{12.60}})      & 37.17 (\textcolor{ForestGreen}{ $\uparrow$ 23.32})               \\
                                          & BM25 (Quad-gram)                  & 26.94 
                                          (\textcolor{ForestGreen}{ $\uparrow$ 12.22})
                                          & \textbf{39.15} (\textcolor{ForestGreen}{ $\uparrow$ \textbf{25.3}})      \\
                                          & BM25 (Penta-gram)                 & 25.81 (\textcolor{ForestGreen}{ $\uparrow$ 11.09})               & 38.61 (\textcolor{ForestGreen}{ $\uparrow$ 24.76})             \\
                                          \hline \hline
\multirow{5}{*}{RR Filtered Docs}          & BM25                              & 12.97 (\textcolor{red}{ $\downarrow$ 1.75})               & 13.05 (\textcolor{red}{ $\downarrow$ 0.80})               \\
                                          & BM25 (Bigram)                     & 21.06 (\textcolor{ForestGreen}{ $\uparrow$ 6.34})               & 24.67 (\textcolor{ForestGreen}{ $\uparrow$ 10.82})               \\
                                          & BM25 (Trigram)                    & 24.97 (\textcolor{ForestGreen}{ $\uparrow$ 10.25})                & 34.22 (\textcolor{ForestGreen}{ $\uparrow$ 20.37})               \\
                                          & BM25 (Quad-gram)                  &         24.90 (\textcolor{ForestGreen}{ $\uparrow$ 10.18})             & 36.77 (\textcolor{ForestGreen}{ $\uparrow$ 22.92})               \\
                                          & BM25 (Penta-gram)                 &             23.72 (\textcolor{ForestGreen}{ $\uparrow$ 9.00})         & 37.72 (\textcolor{ForestGreen}{ $\uparrow$ 23.87})     
             \\ \bottomrule
\end{tabular}
\caption{The table shows the performance comparison (F1 scores in \%, with top K retrieved documents selected using validation set) of the proposed method with the baseline unsupervised methods on the COLIEE-21 \cite{COLIEE2021} and proposed \ILPCR\ benchmark. The numbers in the bracket highlight the performance difference compared to the BM25 (Baseline, Table's first row). $\uparrow$ shows the  increase, and $\downarrow$ shows the drop in performance.}
\label{tab:experiment_results}
\end{table*}

\begin{table*}[!h]
\small
\renewcommand{\arraystretch}{1}
\setlength\tabcolsep{1.1pt}
\centering
\begin{tabular}{cccc}
\toprule
 { Method } & Brief Description & Unsupervised &{ F1 } \\
\midrule  
{ JNLP \cite{JNLP}}  &Top-100,Paragraph,BM25,BERT,Union Score&$\checkmark$& 0.19 \\
{ TR \cite{COLIEE2021}}    &Top-1000 TF-IDF, Xgboost&$\checkmark$& 0.46 \\
{ DSSIR \cite{DoSSIER}} &vanilla BERT& \xmark & 2.79 \\
{ DSSIR \cite{DoSSIER}} &paragraph level BM25, lawDPR& \xmark & 2.72 \\
{ SIAT \cite{COLIEE2021}}  &Top-50 BM25, BERT-Legal&$\checkmark$& 3.00 \\
{ DSSIR \cite{DoSSIER}} &BM25&$\checkmark$& 4.11 \\
{ TLIR \cite{Ma2021RetrievingLC}}  &LMIR, BERT-PLI on paragraphs& \xmark & 4.56 \\
{ NM \cite{Yes-BM25}}    &Vanilla BM25-Segments& $\checkmark$ & 9.37 \\
{ TLIR \cite{Ma2021RetrievingLC}}  &Language Model for IR and paragraph filtering & $\checkmark$ &  19.17 \\
MTFT-BERT \cite{ECIR-22-bert} & Multi-task optimization over $BM25_{optimized}$ & \xmark & 22.05\\ 
\textbf{U-CREAT} & BM25 (Tri-gram) over Events Filtered Docs & $\checkmark$ & \bf{27.32}\\
\bottomrule
\end{tabular}
\caption{The table shows the performance comparison of the proposed method with the existing methods on the COLIEE-21 \cite{COLIEE2021} dataset. The F1 scores (in \%) represent the numbers reported in respective methods. The table highlights a significant performance boost with respect to the current state-of-the-art MTFT-BERT (supervised method trained on COLIEE-21 corpus). 
}
\label{tab:existing_method_comparison}
\end{table*}

\begin{figure*}[t]
  \centering
  \includegraphics[width=0.37\textwidth,trim={0cm 0cm 9.5cm 0cm}, clip]{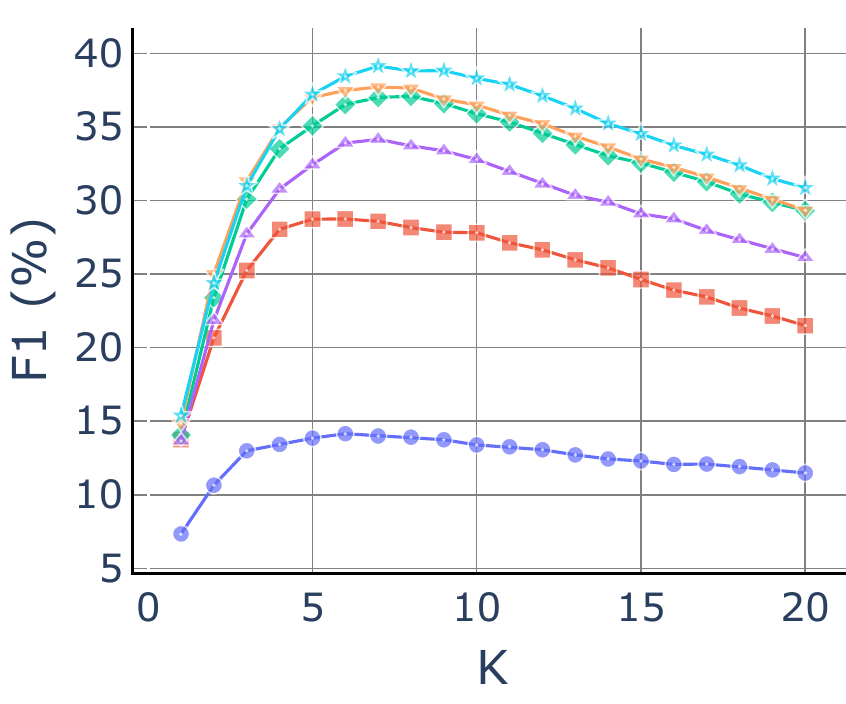}
\includegraphics[width=0.62\textwidth,trim={0cm 0cm 0cm 0cm},clip]{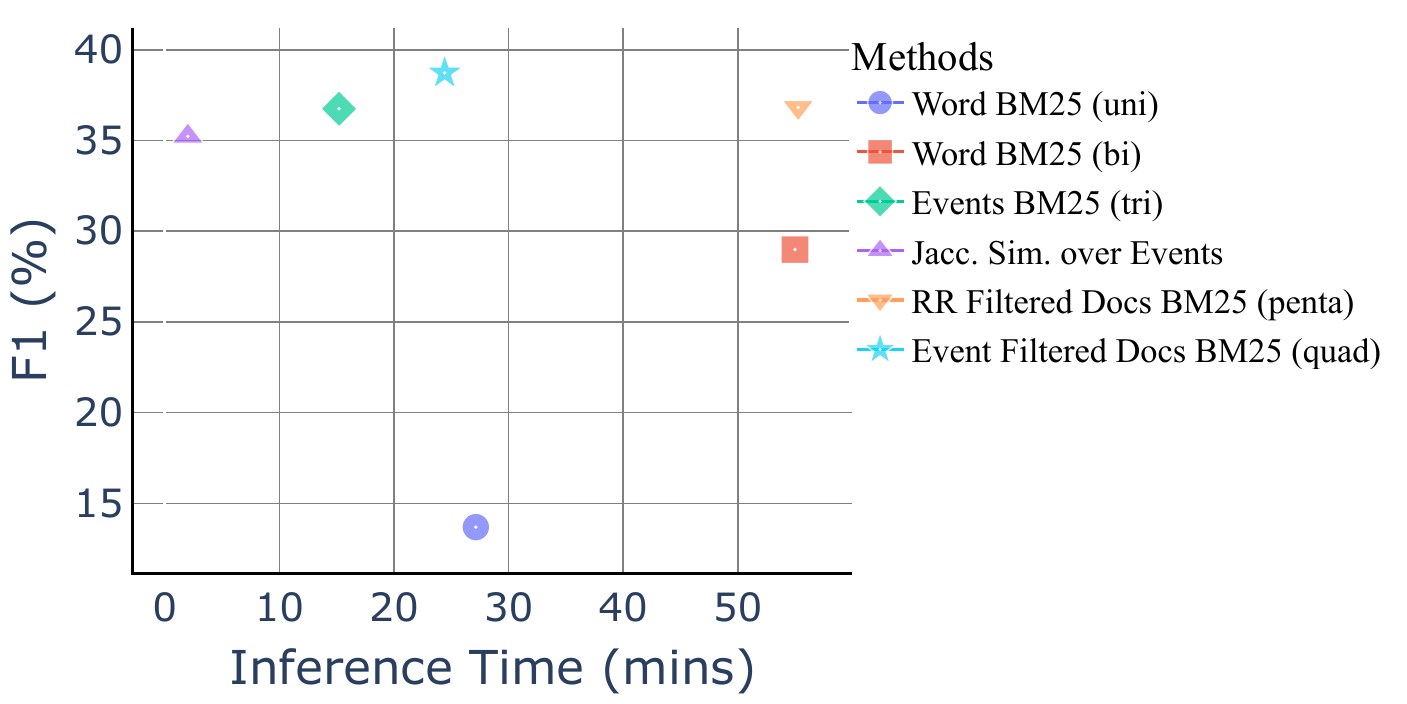}
\caption{The figure on the left shows the performance curves and the right shows inference time vs. performance of various methods. Also see Appendix Table \ref{tab:inference-runtimes}.}
\label{fig:experiment_performance_auc-performance_vs_time_PCR}
\vspace{-5mm}
\end{figure*}

\noindent\textbf{Non-atomic Events:} For this setting, we consider the words (predicates and arguments) that are present in the extracted events $\mathcal{E}^{(\mathcal{Q}_i)}$ and $\mathcal{E}^{(\mathcal{C}_j)} $ separately. This setting removes the event as an atomic unit, and it considers words of each event as an independent unit, i.e., a document is represented only by individual words in the extracted events. We run various variants for BM25 to generate relevance scores. We found that the trigram version of BM25 (the best model for non-atomic events) has a similar performance to the best model for atomic events (BM25). 


\noindent\textbf{Events filtered Docs:} As the primary role of events is to capture the relevance between the query and the candidate doc, for this variant, we select the complete sentences corresponding to the overlapping events $|\mathcal{E}^{(\mathcal{Q}_i)} \cap \mathcal{E}^{(\mathcal{C}_j)}|$. For example, if a common event $e^{\mathcal{Q}_i}_p$ emanates from sentences $\mathcal{S}_t$ and $\mathcal{S}_v$ in the query and candidate document, respectively, we consider the sentence  $\mathcal{S}_t$ from the query and $\mathcal{S}_v$ from the candidate. Selecting sentences for each overlapping event results in sentences selected for every doc. We refer to this updated version of the doc as the events filtered doc and use this new version for classical retrieval methods like BM25. We observe that this setting further improves the retrieval scores by $2.62$ in \ILPCR\ and $3.19$ in COLIEE'21, compared to the best non-atomic event-based methods. Overall, this setting outperforms all the other methods for both datasets and shows a performance boost of $25.3$ F1 score in \ILPCR\ and $12.6$ F1 score in COLIEE'21 compared to the standard BM-25 baseline.

\noindent \textbf{Event Embeddings:} We also tried models based on event embeddings obtained by composing embeddings of predicates and arguments, e.g., via transformer models or deep NNs \cite{modi-2016-event, modi-titov-2014-inducing}; however, these approaches gave a worse performance than vanilla transformer based approaches. Moreover, these approaches have an extra overhead of training (and learning) event embeddings.  

\noindent\textbf{Rhetorical Roles Filtered Docs:}
In the legal domain, Rhetorical Roles (RR) \cite{malik-rr-2021,kalamkar-etal-2022-corpus} have been introduced to segment a document into semantically coherent textual units corresponding to 7 main rhetorical roles: Facts, Arguments, Statues, Ruling, Precedents, Ratio, and Judgment. For more details, please refer to \citet{malik-rr-2021}. The main idea is to label each sentence in the legal document with one of the rhetorical roles. For RR, we used the pre-trained transformer-based model utilizing multi-task learning provided by \citet{malik-rr-2021} to predict sentence-level labels for legal documents in COLIEE21 and \ILPCR. We used some specific RR labels (that capture relevance as per legal experts) to filter out sentences from a query (RRs used: facts, argument, ratio) and candidates (RRs used: facts, argument, ratio, and judgment). Using all RRs labels gave a worse performance, possibly due to the introduced noise. The filtered query and candidate documents are then used for BM-25-based baselines. Table \ref{tab:experiment_results} shows that a pre-filtering step done using a pre-trained RR model is a strong retrieval method and provides a significant performance boost (increase of $24.97$ in COLIEE'21 and $37.72$ in the case of \ILPCR\ ). However, the events-based filtering methods remain the outperforming model ($27.32$ increase in COLIEE'21 and $39.15$ boost in F1 score in \ILPCR). However, in the case of RR, inference time in the case of quad-gram and penta-gram increases drastically, making them impractical (\S\ref{sec:analysis}). RR-based models have lesser improvement on COLIEE'21 as the pre-trained RR model \cite{malik-rr-2021} used for labeling is trained on Indian legal documents.  

\subsection{Comparison with Existing Methods}\label{sec:existing-results}

For a fair comparison with the existing methods, we compare the proposed event-based approaches with the state-of-the-art methods for the COLIEE'21 benchmark. A recent supervised retrieval approach by \citet{ECIR-22-bert} uses a multitasking framework to improve upon the optimized BM25 retrieval scores. To the best of our knowledge, this approach is the current state-of-the-art method for the COLIEE-21 document retrieval task. Table \ref{tab:existing_method_comparison} shows the F1 scores obtained by multiple methods, as given in \cite{COLIEE2021}. The proposed event-based methods outperform the existing approaches by a significant margin highlighting the effectiveness of events in legal document retrieval. A noteworthy point here is that the event-based techniques are completely unsupervised, making them more applicable to current systems without corpus-specific training. Moreover, these approaches generalize well over legal documents in different legal systems, as shown using two different legal system datasets.


\begin{figure*}[t]
  \centering
  \includegraphics[width=0.37\textwidth,trim={0cm 0cm 9.5cm 0cm}, clip]{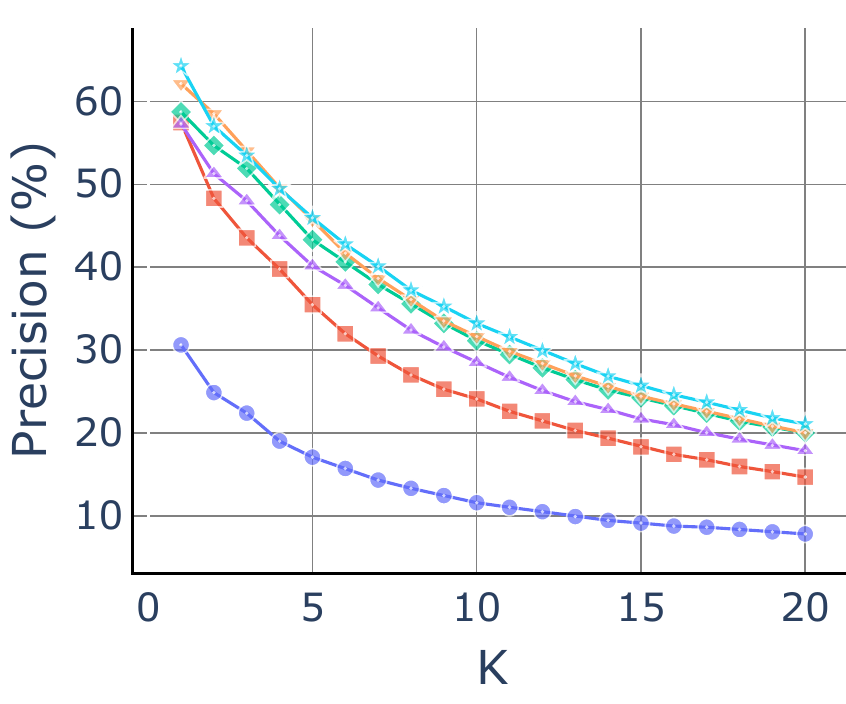}
\includegraphics[width=0.62\textwidth,trim={0cm 0cm 0cm 0cm},clip]{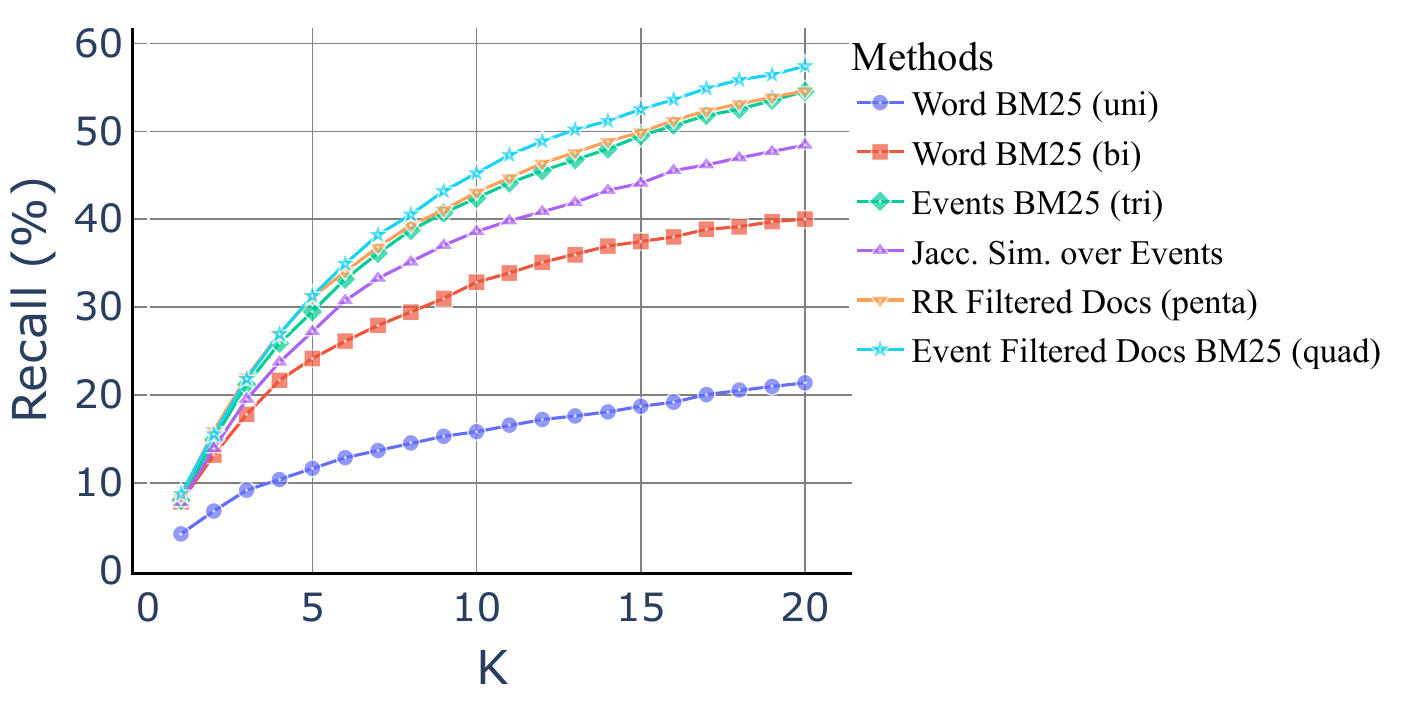}
\caption{The precision and recall curves at different K (top retrieved documents) values used for retrieval.}
\label{fig:precision_recall_plot}
\vspace{-5mm}
\end{figure*}

\subsection{Analysis}\label{sec:analysis}
\noindent\textbf{Variation with K:} To provide a detailed insight into the performance of various methods, we also show the F1 score at different K values (top retrieved documents) on \ILPCR. Figure \ref{fig:experiment_performance_auc-performance_vs_time_PCR} (left side) highlights the improvements in the F1 curves obtained by event methods compared to the popularly used BM25 baselines. The performance peaks for K = 3 to 7, this is similar to what has been observed on the COLIEE dataset \cite{COLIEE2021}. We show the variation of Precision and Recall scores with the value of K in Figure \ref{fig:precision_recall_plot}. As can be observed (and is expected based on the evaluation metric definition) for each of the models, precision falls and recall improves with increasing K values, resulting in the hump shape in Fig. \ref{fig:experiment_performance_auc-performance_vs_time_PCR}. The Precision, Recall, and F1 scores corresponding to best K are tabulated in Appendix Table \ref{tab:precision-recall-results}.  


\noindent\textbf{Inference Time:} An important property of a retrieval algorithm often not stressed by existing methods is inference time. For a retrieval system to be adaptable to industrial solutions, it is not only the retrieval efficiency but also the inference time required by the system. We compare inference times of various methods to provide a more transparent insight. We use the queries in the entire test split (237 query documents) of the \ILPCR\ corpus to calculate inference time. We benchmark the relevance score generation time for all the queries on a single core of an Intel(R) Xeon(R) Silver 4210R CPU @ 2.40GHz processor. We compute the event extraction time along with the relevance score generation time for the proposed event-based methods. Figure \ref{fig:experiment_performance_auc-performance_vs_time_PCR} (right side) shows the inference vs. performance comparison for the prominent methods (also see exact numbers in App. Table \ref{tab:inference-runtimes}). 

The inference time for the different models varies greatly, the Jaccard Similarity over Events (IOU) stands out with the fastest time of $2$ minutes, while the Word BM25 (bigram) model has the longest inference time of $55$ minutes. The Events BM25 (trigram) model has a much faster inference time of $15.2$ minutes, which is approximately $50$\% faster than the Word BM25 (unigram) model. The Event Filtered Docs BM25 (quadgram) model also has a relatively fast inference time of $24.42$ minutes, which is about $10\%$ faster than the Word BM25 (unigram) model. Overall, the proposed Event Filtered Docs BM25 (quadgram) has a relatively fast inference time of $24.42$ minutes compared to the other models and represents a significant improvement in performance. This time is about $10\%$ faster than the Word BM25 (unigram) model and significantly faster than the Word BM25 (bigram) model, which has the longest inference time of $55$ minutes. In the F1 score, the Event Filtered Docs BM25 (quadgram) model also outperforms the Jaccard Similarity over Events (IOU) model, which has the quickest time at 2 minutes, and the Events BM25 (trigram) model, which has a time of $15.2$ minutes. The proposed model stands out as a strong performer in terms of inference time and F1, providing a significant improvement in performance compared to the other models.

In terms of inference time, the retrieval method based on Jaccard similarity shows a significant performance boost along with a significant improvement in the F1 score. Overall, the increase in document size results in a longer inference time in BM25-based methods. Moreover, going from unigram to bigram also results in a considerable increase in inference time, making the word-based BM25 bigram ineffective for real-time retrieval systems. The inference time results for event-based methods highlight the effectiveness both in terms of inference time and retrieval efficiency. 

A noteworthy trend in the current deep learning-based supervised methods in legal document retrieval is the use of BM25 as a pre-filtering step \cite{DESIRES,JNLP,UIRPC,Shao2020BERTPLIMP,ECIR-22-bert}. The scores obtained from a word-based BM25 provide a strong pre-filtering, enabling the re-ranking-based algorithm to improve the scores over the top-K\% retrieved documents. This re-ranking setting for inference on a deployable system would require BM25 inference time and  deep model inference time to generate the retrieval scores. In contrast, the proposed event-based approaches lead to a much faster inference time and improved retrieval performance. It would facilitate the current research on supervised retrieval methods as well.

\noindent\textbf{Other Observations:} We also experimented with another version of the corpus where we removed the sentences containing the citation to prevent the model from exploiting any neighboring information. The results are shown in the Appendix Table \ref{tab:results-without-citations}, there is a slight drop in performance; however the overall trends (as in Table \ref{tab:experiment_results}) remain the same. 

\subsection{Discussion}
An important point to note is that the PCR task has inherent limitations; the relevant cases are considered based on official citations as ground truth. However, there might be cases that were not mentioned by the judge (document writer) due to subjectivity involved in the common-law system; finding correct annotation for relevance is always a challenge for a domain like legal, where the number of documents is enormous. 

\section{Conclusion}\label{sec:conclusion}

In this paper, we proposed a new large dataset (\ILPCR) for Prior Case Retrieval and the U-CREAT pipeline for performing event-based retrieval for legal documents. We ran a battery of experiments with different types of models to show that event-based methods have better performance and much better inference times (and hence amenable to production settings) compared to existing unsupervised approaches and some of the supervised approaches (e.g., $\sim 5.27$ F1 score improvement on COLIEE) on two completely different datasets. In the future, we plan to combine event-based methods with supervised techniques such as contrastive learning to develop more efficient models.

\section*{Limitations}\label{sec:limitation}
In this paper, we propose a simple model for prior case retrieval. As shown in experiments and results, the models could improve and score better. There is a big room for improvement. All the previously proposed approaches for PCR have calculated relevance as some form of lexical/semantic similarity between a case and its citations. However, cited case relevance may sometimes differ from lexical/semantic similarity. Modeling the document in terms of events only partially addresses this. Consequently, what is required is the inclusion of more legal information. We made an attempt towards that via experiments using Rhetorical Roles. Similarly, one could use the information coming via statutes and laws since similar cases are likely to invoke similar statutes. Another approach is learning representations using contrastive models that score relevant cases higher than non-relevant ones. In the future, we plan to investigate these approaches to improve the task of PCR. 


This paper considers a simple structure for an event as a tuple of predicates and arguments. However, more sophisticated formulations are possible, as outlined in the survey/tutorial by \citet{chen-etal-2021-event}. Moreover, we are taking events in isolation and ignoring the sequential nature of events that help to form narratives. In the future, we would like to develop a model that captures a more sophisticated structure and sequential nature of events in the case. Though we covered an extensive set of experiments for the proposed event-based matching technique, many more combinations can be experimented with to understand the role of events in legal documents. This unique finding of events missing from the legal literature would facilitate exploring new directions in the legal domain.

In this paper, we evaluated only two datasets as we could not find any publicly available PCR datasets. However, in the future, if we can find more PCR datasets, we would like to evaluate them to see if the trends generalize over other legal corpora.

\section*{Ethical Concerns}
This paper proposes a system for retrieving (recommending) relevant documents. The system is not involved in any decision-making process. The motivation for proposing the system is to augment legal experts rather than replace them. Moreover, for training the system, we used publicly available legal documents. We took steps to normalize documents concerning named entities to prevent a model from developing any known biases. To the best of our knowledge, we addressed any biases that the model might learn from the data. 

\bibliography{references}
\bibliographystyle{acl-natbib.bst}

\clearpage
\newpage
\appendix
\section*{Appendix}

\section{Evaluation Metric Definition} \label{app:eval-metric}

\begin{align*}
    &\text{Precision} = \frac{(\text{\# correctly retrieved cases $\forall$ queries})}{(\text{\# retrieved cases $\forall$ queries})},\\
    &\text{Recall} = \frac{(\text{\# correctly retrieved cases $\forall$ queries})}{(\text{\# relevant cases $\forall$ queries})}, \\
    &\text{F1} = \frac{(\text{2 x Precision x Recall})}{(\text{Precision + Recall})}
\end{align*}

\section{Hyper-Parameters} \label{app:hyper-params}

\noindent\textbf{Transformer-Based Models}: We train the standard BERT and DistilBERT models using PyTorch and HuggingFace library-based \cite{wolf2019huggingface} implementations for 6 epochs with a batch size of 32 and AdamW \cite{adamW} optimizer with a learning rate of $1\times 10^{-5}$.

\noindent\textbf{Sentence Transformer-Based Models}: We use a batch size of 512 and fine-tune the models for 20 epochs with Adam \cite{Kingma2015AdamAM} of learning rate $5\times 10^{-5}$ 

\section{SBERT Fine Tuning Strategy} \label{app:simcse}

SBERT is finetuned using SimCSE \cite{simcse} based checkpoints present in SBERT package \cite{reimers2019sentence}, due to the unavailability of annotated similar sentence pairs present for the datasets, SimCSE is trained in unsupervised manner by predicting the input sentence itself using dropout for noisy representation of the sentence.

\section{Precision and Recall Scores}

Table \ref{tab:precision-recall-results} shows the Precision, Recall and F1 scores for various models in given in the main paper. Table \ref{tab:results-without-citations} shows the Precision, Recall and F1 scores for various models on the version of \ILPCR\ without citation sentences. 

\section{Inference Time of Models}

Table \ref{tab:inference-runtimes} shows the inference time for algorithms shown in Fig \ref{fig:experiment_performance_auc-performance_vs_time_PCR}.

\begin{table}[h]
\small
\renewcommand{\arraystretch}{0.7}
\setlength\tabcolsep{2pt}
\hspace{0.2cm}
\centering
\begin{tabular}{cc}
\toprule
\textbf{Algorithm}              & \textbf{Inference Time (mins)} \\ \midrule
Word BM25 (unigram) & 27.14                \\
Word BM25 (bigram) & 55.00            \\
Events BM25 (trigram) & 15.20               \\
Jaccard sim. over events & 2.00               \\
RR filtered BM25 (penta) & 55.27                \\
Events filtered BM25 (quad) & 24.42               \\ 
\bottomrule
\end{tabular}
\caption{Inference Times for various models.}
\label{tab:inference-runtimes}
\end{table}

\begin{table*}[]
\setlength\tabcolsep{6pt}
\centering
\begin{tabular}{cccccc}
\toprule 
\multicolumn{2}{c}{\textbf{Model}}                                                       &
\textbf{K} &
\textbf{Precision} &
\textbf{Recall} &
\textbf{F1}
\\ \midrule
\multirow{2}{*}{Word Level}                & BM25                                             & 5 &   17.11    & 11.64 & 13.85        \\
                                           & BM25 (Bigram)                                   & 7 &  29.30  & 27.91 & 28.59           \\
                                           \midrule
\multirow{6}{*}{\begin{tabular}{@{}c@{}}Segmented-Doc  \\ Transformer \\(full document) \end{tabular}}     
& BERT               & 6 &   10.28   & 8.40 & 9.24             \\
                                          & BERT (finetuned)    & 6 &    8.79    &        7.18 & 7.90       \\
                                          & DistilBERT                       & 7 &         17.02         & 16.21  & 16.61             \\
                                          & DistilBERT (finetuned)            & 5 &       9.70         &    6.60  & 7.86         \\
                                          & InCaseLawBERT     & 11 &        3.02             & 4.52  & 3.62             \\
                                          & InLegalBERT     & 12 &          6.10            & 9.96 & 7.56                \\
                                          \midrule
\multirow{6}{*}{\begin{tabular}{@{}c@{}} Transformer \\(top 512 tokens) \end{tabular}}
& BERT              & 20 &        0.38           & 1.04 & 0.56                \\
                                          & BERT (finetuned)   & 15 &        0.65          &    1.33  & 0.87        \\
                                          & DistilBERT                    & 20 &         0.34           & 0.93   & 0.50        \\
                                          & DistilBERT (finetuned)         & 20 &         0.51          & 1.39   & 0.75            \\
                                          & InCaseLawBERT     & 20 &           0.51         & 1.39 & 0.75               \\
                                          & InLegalBERT     & 19 &          0.49            & 1.27
                                          & 0.71 \\
                                          \midrule
\multirow{4}{*}{\begin{tabular}{@{}c@{}} Sentence \\ Transformer \\(SBERT) \end{tabular}}
& BERT        & 5 & 7.35 & 4.98 & 5.94  \\
& DistilRoBERTa & 4 & 5.56 & 3.01 &  3.91  \\
& BERT (finetuned) & 5 &  7.44 & 5.04 & 6.01\\
& DistilRoBERTa (finetuned)& 7 & 2.20 & 2.08 & 2.14 \\
                                          \hline \hline                                          
\multirow{4}{*}{Atomic Events} & Jaccard Similarity & 7 & 35.12 & 33.28 & 34.17  \\ & BM25                                           & 7 &    37.69    & 35.90 & 36.77       \\
                                           & BM25 (Bigram)                         & 6 &     35.39    & 28.89 & 31.81        \\
                                           & BM25 (Trigram)                                & 6 &     30.71   &  25.07 & 27.61       \\
                                           
                                           \midrule
\multirow{5}{*}{Non-atomic Events}              & BM25                                           & 6 &   13.33  & 10.89   & 11.99        \\
                                           & BM25 (Bigram)                                 & 7 &    33.07   & 31.50   & 32.27       \\
                                           & BM25 (Trigram)                                & 6 &   40.64    & 33.18  & 36.53  \\
                                           & BM25 (Quad-gram)                              & 7 &    35.62   & 33.93  & 34.76         \\
                                           & BM25 (Penta-gram)                              & 6 &    37.30    & 30.46  & 33.54          \\
                                           \midrule

\multirow{5}{*}{Events Filtered Docs}      & BM25                                   & 5 &    24.26   & 16.50   & 19.64       \\
                                           & BM25 (Bigram)                                  & 6 &    33.69   & 27.50  & 30.28           \\
                                           & BM25 (Trigram)                       & 6 &    41.35   & 33.76 & 37.17        \\
                                           & BM25 (Quad-gram)                               & 7 &  40.12   & 38.22  & 39.15  \\
                                           & BM25 (Penta-gram)                              & 7 &     39.57     & 37.70 & 38.61       \\
             \hline \hline
\multirow{5}{*}{RR Filtered Docs}          & BM25                    & 7 &          13.37                & 12.74   & 13.05            \\
                                          & BM25 (Bigram)                 & 7 &           25.29        & 24.09   & 24.67            \\
                                          & BM25 (Trigram)            & 7 &          35.08             & 33.41    
                                          & 34.22 \\
                                          & BM25 (Quad-gram)            & 7 &          37.69           & 35.90     &  36.77          \\
                                          & BM25 (Penta-gram)            & 7 &     38.66        &  36.83  & 37.72     
             \\ \bottomrule
\end{tabular}

\caption{The table shows the K values, Precision, Recall and F1 scores for each model.}
\label{tab:precision-recall-results}
\end{table*}

\begin{table*}[]
\setlength\tabcolsep{8pt}
\centering
\begin{tabular}{cccc}
\toprule 
\multicolumn{2}{c}{\textbf{Model}}                                                       & \textbf{\ILPCR} & 
{\begin{tabular}{@{}c@{}}\textbf{\ILPCR\textsubscript{$\neg$sent}} \\ (without citation sents.) \end{tabular}} \\ \midrule
\multirow{2}{*}{Word Level}                & BM25                                             & 13.85      & 13.23         \\
                                           & BM25 (Bigram)                                    & 28.59    & 27.52           \\
                                           \midrule
\multirow{6}{*}{\begin{tabular}{@{}c@{}}Segmented-Doc  \\ Transformer \\(full document) \end{tabular}}     
& BERT                 & 9.24       & 9.58              \\
                                          & BERT (finetuned)     & 7.90               &        8.41        \\
                                          & DistilBERT                        & 16.61                & 17.58               \\
                                          & DistilBERT (finetuned)            & 7.86               &    8.21            \\
                                          & InCaseLawBERT         & 3.62                & 3.25               \\
                                          & InLegalBERT           & 7.56               & 7.96               \\
                                          \midrule
\multirow{6}{*}{\begin{tabular}{@{}c@{}} Transformer \\(top 512 tokens) \end{tabular}}
& BERT                 & 0.56               & 0.36                \\
                                          & BERT (finetuned)     & 0.87              &    0.67           \\
                                          & DistilBERT                        & 0.50             & 0.52               \\
                                          & DistilBERT (finetuned)            & 0.75             & 0.68               \\
                                          & InCaseLawBERT         & 0.75             & 0.68               \\
                                          & InLegalBERT           & 0.71              & 0.68               \\
                                          \midrule
\multirow{4}{*}{\begin{tabular}{@{}c@{}} Sentence \\ Transformer \\(SBERT) \end{tabular}}
& BERT         & 5.94 & 4.73  \\
& DistilRoBERTa &  3.91 & 2.94  \\
& BERT (finetuned) & 6.01 & 5.01 \\
& DistilRoBERTa (finetuned) & 2.14  & 1.01 \\
                                          \hline \hline                                          
\multirow{4}{*}{Atomic Events} & Jaccard Similarity & 34.17 & 32.38  \\ & BM25                                             & 36.77       & 35.26        \\
                                           & BM25 (Bigram)                           & 31.81      & 30.96         \\
                                           & BM25 (Trigram)                                   & 27.61    &  26.59       \\
                                           
                                           \midrule
\multirow{5}{*}{Non-atomic Events}              & BM25                                             & 11.99    & 11.99           \\
                                           & BM25 (Bigram)                                    & 32.27   & 31.91 \\
                                           & BM25 (Trigram)                                   & 36.53  & 36.02    \\
                                           & BM25 (Quad-gram)                                 & 34.76    & 33.75           \\
                                           & BM25 (Penta-gram)                                & 33.54     & 32.38          \\
                                           \midrule

\multirow{5}{*}{Events Filtered Docs}      & BM25                                    & 19.64     & 19.78          \\
                                           & BM25 (Bigram)                                    & 30.28    & 30.35           \\
                                           & BM25 (Trigram)                          & 37.17      & 36.40         \\
                                           & BM25 (Quad-gram)                                 & 39.15  & 38.32    \\
                                           & BM25 (Penta-gram)                                & 38.61        & 37.66       \\
             \hline \hline
\multirow{5}{*}{RR Filtered Docs}          & BM25                              & 13.05               & 13.65               \\
                                          & BM25 (Bigram)                     & 24.67               & 24.80              \\
                                          & BM25 (Trigram)                    & 34.22                & 33.15               \\
                                          & BM25 (Quad-gram)                  & 36.77              & 36.77               \\
                                          & BM25 (Penta-gram)                 & 37.72         & 36.93     
             \\ \bottomrule
\end{tabular}

\caption{The table shows the performance comparison (F1 scores in \%) of the proposed method with the baseline unsupervised methods on the COLIEE-21 \cite{COLIEE2021}, the \textbf{IL-PCR} benchmark and the dataset with sentences having the citation removed: \textbf{IL-PCR\textsubscript{$\neg$sent}})}
\label{tab:results-without-citations}
\end{table*}

\end{document}